# THE USE OF DATA FROM INFORMATION SYSTEMS IN COURT PROCEEDINGS

Dobromira Bankova and Vladimir Dimitrov

**Absract:** This paper examines data in the context of how the judiciary collects, analyses, and evaluates it as evidence, based on examples from current judicial practice in Bulgaria and within the context of the new substantive legal regulations. It explores the legal and practical challenges related to the use of data sets as evidence in court proceedings through the analysis of specific cases. In light of the new regulatory framework, the research points out that the analytical perspective should shift from "evidence as an information unit" towards "evidence as a behavioural algorithm", requiring not only technological tools but also a methodological shift and adequate preparation for collecting and assessing aggregated digital evidence.

## 1. Introduction

This article aims to explore the emerging legal, procedural, and technological challenges associated with the use of data from information systems as evidence in judicial proceedings. In the context of digital transformation and the adoption of the new European regulatory framework for data, the judiciary is increasingly required to adapt both its approach towards electronic evidence and its understanding of the admissibility, reliability, and explainability of datasets.

Considering existing practical realities and current regulatory mechanisms in Bulgaria, the article argues for an objective necessity to reconsider how the judiciary handles digital evidence concerning datasets. This research uses the term digital evidence as highlighting the technological characteristic of digital data as electronic evidence. In particular, it is essential to recognize that data can and should be presented and analyzed as algorithms revealing comprehensive behavioral patterns, rather than merely as isolated informational units. This, in turn, necessitates establishing clear procedural rules governing their use, verification, and contestation within judicial proceedings.

The article is structured into six sections, each addressing key aspects of the discussed topic. The second section explores the legal concept of data as an autonomous object distinct from databases, emphasizing the definition and legal nature of data under the new regulatory framework. Section three analyzes specific examples from judicial practice, illustrating real and current challenges in presenting and assessing data as evidence before courts. Section four focuses on regulatory issues related to the Data Act, examining its impact on procedural law, the right of defense, and the role of the judiciary as a guarantor of legality. This section also addresses issues related to the use of artificial intelligence in judicial proceedings, including transparency, explainability, and controllability requirements for automated decision, as a distinct aspect within the broader discussion regarding datasets. Section five identifies the main practical and procedural challenges. It examines how data should be processed in judicial proceedings, whether as isolated informational units or through analysis of behavioral patterns derived from datasets. It also discusses issues related to data collection and access, procedural safeguards, the need for expert capacity and the necessary technological infrastructure. The final, sixth section outlines the main conclusions and recommendations for the more effective use of data as evidence in judicial proceedings. It summarizes the key findings and attempts to identify possible directions for future development.

By combining legal analysis, practical case studies, and regulatory context, the study aims to provide a comprehensive understanding of the transformation required by the judicial system to meet the demands of the digital age and ensure fairness in an information-rich reality.

## 2. Data as an independent legal object: definitions, distinctions, and relevance to the evidentiary process

One of the key aspects of the contemporary debate on digital evidence is the question of what exactly should be understood by the term "data" in a legal context. Until recently, the legal framework in the European Union focused predominantly on databases as objects of protection, particularly in the field of intellectual property rights protection. In the case of databases, individual informational units were considered with regard to their organizational structure, which was subject to creation, use, and legal protection through rights similar to copyright. As a result, legal enforcement and judicial practice seemingly did not assign independent significance to unstructured data that did not fall within the scope of the established databases.

With the adoption of Regulation (EU) 2023/2854 (the Data Act) [1], this framework is undergoing transformation. According to Article 2 (1) of the Regulation, "data" means any digital representation of acts, facts, or information, and any compilation of such representations that can be stored or processed electronically. This definition is key for several reasons. It does not recognize a distinction between structured and unstructured data. Unlike databases, which imply organization, logical structure, and selection, data under the Regulation can exist autonomously—for example, in the form of log files, video recordings, sensor outputs, automatically generated reports, and many others. Data is no longer treated merely as an element within a larger structure but as an independent object of rights and obligations. According to us, this understanding of the substantive legal nature of data influences procedural behavior when data is used as evidence in judicial proceedings, even when it does not exist in the usual form of documents or materialized declarations.

Typically, individual informational units are combined into a system that might not necessarily be produced or structured. To distinguish this type of data combination from the term "databases," in the context of this research, we use the term "datasets." **Datasets refer to groups of digital elements, structured or unstructured, interconnected by an objective criterion such as time and/or source, and/or subject, and/or physical location, and/or origin, which can be jointly analyzed to draw conclusions about behaviors, processes, or events.** This includes, but is not limited to: log files of user activity, sequences of communications, digital traces of electronic transactions, related images, audio or video files, geolocation and telemetry data, among others. It should be emphasized that "datasets" are not synonymous with the concept of "big data." The latter implies not only the existence of large data volumes but also high variability and velocity in data generation and updating, requiring specialized analytical technologies, and represents a distinct concept in a substantive legal context. In judicial proceedings, in our view, it is more common to encounter datasets that do not reach the thresholds of "big data" and thus fall outside its official definition, yet still require interpretation through a comprehensive, aggregated approach.

For precision, it should be noted that this research also analyzes cases involving big data in the context of artificial intelligence, where datasets acquire the characteristics of big data, particularly when containing dynamically generated input data or when used in training models. Nevertheless, for the sake of definitional accuracy, we distinguish the term "datasets," as used in this research, from the established concept of big data.

The legal recognition of datasets as evidentially relevant objects, regardless of their form and structure, constitutes a conceptual breakthrough. This opens the door to the utilization of previously overlooked evidentiary resources, simultaneously requiring new standards of admissibility, authenticity, sufficiency, integrity, and expert verifiability. This new way of thinking establishes a connection between data regulation, procedural law, and the necessity for a structured analytical framework for digital information as a legally relevant fact.

### 3. Case-based analysis of practical applications: examples from judicial practice

After clarifying the respective definitions and their significance in the previous section, this part of the research examines three specific cases from current Bulgarian judicial practice, illustrating how data from information systems is used and analyzed as evidence by the courts. The selection of these examples has been made because the conclusions are based on current reasoning from judicial acts, involve various types of legal relationships, and consequently raise different questions regarding the admissibility, reliability, and interpretation of datasets in judicial proceedings. What they have in common is that each case, in its own way, highlights the necessity for courts to adopt a new approach towards datasets—considering them not merely as isolated informational units but as part of an algorithm establishing broader behavioral patterns.

The first case concerns the use of Special Investigative Means (SIM). The application of SIM typically generates large volumes of audio data, which are subsequently formalized into written transcripts. The problem arises when the authenticity of the transcripts is challenged, requiring the court to compare the transcript with the original audio recording. In a broader sense, this raises issues related to the integrity of evidence and procedural safeguards concerning the processing of large volumes of audio data. When the accused contests the authenticity or accuracy of the transcript, the court is obligated to request the original audio recording, allow the parties to listen to it, compare it with the transcript, and, if discrepancies are found, appoint phonographic expertise. Any inaccuracy may constitute a procedural violation and may lead to the exclusion of evidence under Art. 105 of the Bulgarian Criminal Procedure Code [6] ("evidence obtained in violation of the law"). If the disputed evidence is crucial for the prosecution's case, such a procedural deficiency could compromise the entire prosecutorial framework.

When analyzing digital evidence as datasets, the court must adopt a model-based rather than a fragmented approach. This includes examining the complete chronology of communications (timestamps, frequency, participants), contextual analysis (i.e., when and under what circumstances the statement was made), communication dynamics, contradictions among statements over time (e.g., whether a later conversation objectively disproves or merely reinterprets an earlier one), as well as statistical frequency and repetition,for instance, whether a given statement (e.g., "we will take the money") was made only once or multiple times in various forms. Apart from the procedural validity of evidence, the court must also assess whether it sufficiently proves the assertion.

These considerations should be viewed in the context of the evolving regulatory landscape, including Directive 2014/41/EU and the emerging EU initiative on digital evidence [7]. This raises the issue of presenting, analyzing, and evaluating data as a sequence, as an integrated process rather than as an isolated moment or fragmented informational unit. This specific case also highlights the question of whether an electronic dataset is sufficient in scope, meaning it contains logically related, technically reliable, and legally relevant data that collectively enable the reconstruction and verification of specific human behavior.

The second case involves a civil dispute between a customer and a bank, where the customer claims to have suffered financial damage due to fraud carried out through a classic phishing attack—allegedly facilitated by a breach in the bank's information security. In the framework of Sofia District Court Judgment of 04.03.2025 in civil case No. 20231110148332, year 2023 [8], the court analyzed several categories of digital evidence: log files from mobile applications (SMS, activation timestamps);

data from authorization systems (Google Wallet/Apple Pay); data showing activation of the bank's mobile app on an unknown device; conclusions of a court-appointed technical expert explaining the transaction mechanism; SMS messages containing activation instructions and warnings; telemetry data (sequence and timing of actions); and witness testimonies regarding the website used and the client's actions. The bank provided internal identification and authorization data, including device data and sequences of verification codes.

Some of this evidence is digital, generated on various devices at different times and within different information systems, which contributes to the complexity of evidence-gathering in cases of so-called sophisticated phishing. In this context, the court distinguished the necessity to establish objective facts of a technical nature—particularly whether the bank's electronic system functioned as intended, i.e., as an integrated network of devices and applications, including phones, computers, and banking platforms. An essential element for objectively establishing facts was the chronological and systematic tracking of the client's actions across different platforms, locations, and devices, to determine whether the payer's actions were intentional, negligent, or grossly negligent.

The court allowed a comprehensive analysis of digital interactions (sequence, timing, behavioral patterns), admitting log files and automated messages as admissible and relevant evidence, and evaluated the correspondence between received instructions and user actions as an objective criterion for determining the client's culpability. The approach was integrative, linking technical and behavioral dimensions by analyzing data as a chain of actions and behavioral indicators rather than as an isolated transaction. In the context of this case, and after establishing the existence of a sophisticated phishing attack, the court concluded, according to the applicable legal framework, that it was not necessary to identify the individual who breached the banking system or directly caused the harm in the specific case.

The third case concerns the decision in Constitutional case No. 33/2024 [9], in which the Constitutional Court of the Republic of Bulgaria, within the context of a dispute over the legality of an election, examined data from electronic voting machines, including electronic protocols, device memory, and video recordings. Several key types of data were presented to the court: protocols from sectional election commissions, used to verify consistency between reported results and actual votes; flash memory and device storage containing machine voting data; video recordings from polling stations, accepted as evidence when available, particularly to observe the vote-counting process; expert analyses based on processed data from 1,768 polling stations; and metadata, including timestamps for individual votes. Clearly, some of this data constitutes digital evidence of diverse origins, objective media, and characteristics, yet it is collected and analyzed collectively, alongside written evidence.

Apart from answering the questions posed to the court, the Constitutional Court's decision also provides an additional response of substantial importance regarding the standards for admissibility and reliability of digital data. The Constitutional Court accepted only officially submitted and verifiable digital data, such as those provided by the Central Election Commission (CEC), and did not treat the metadata from the voting machines as directly related to the election results, since the Election Code does not recognize them as an official reporting mechanism. The court emphasized the regulatory requirements for structured data that permit judicial verification, as well as expert analysis based on clearly defined tasks and subject matter. The analysis reveals that the court accepts digital evidence when supported by legal regulation and methodologically sound expert reports. As procedural safeguards when working with such datasets, the court emphasizes the importance of traceability and contextualization, including clear identification of the timestamp, scope, validity, and origin. Isolated digital data without context or verification, such as fragments from flash memory without corresponding protocols, were not accepted.

The examples presented in this section were chosen for their relevance and significance, as they further demonstrate the new procedural safeguards required for analyzing data as digital evidence. They also illustrate how judicial practice is gradually developing an understanding of the complexity and contextual nature of datasets. In this regard, the selected cases support the authors' central argument that it is essential to establish a procedural framework that accounts for the specific characteristics of contemporary information sources in the context of using datasets as evidence in court.

## 4. Regulatory and technological context: new frameworks for handling data in judicial proceeding

The previous section examined specific cases from recent judicial practice illustrating the need for a new approach towards digital evidence. In this section, we examine the existing regulatory and technological framework within the scope of Bulgarian procedural law. The emergence of new technologies and the growing use of datasets and big data highlight the importance of recent European regulations on data and artificial intelligence, which, in turn, reinforce the argument for rethinking the evidentiary process in contemporary judicial proceedings.

This section is structured into three consecutive parts. The first part (4.1) analyzes Regulation (EU) 2023/2854, the Data Act [1], outlining how its provisions, although not procedural in nature, establish a new legal basis for accessing and using digital data in judicial proceedings. Special emphasis is placed on the role of the judiciary as a guarantor of information rights and on the necessity of reforming the evidentiary framework. The second part (4.2) examines what it means to interpret data not merely as isolated informational units but as behavioral patterns that must be extracted and analyzed in their entirety. This shift in perspective requires adaptation at both procedural and expert levels, particularly in cases involving digital fraud, market infringements, or other legal issues requiring comprehensive behavioral reconstruction. The third part (4.3) focuses on the use of artificial intelligence in the context of data in judicial proceedings and raises the question of whether and how the judiciary can exercise effective oversight over automated analyses and decisions. Concepts such as algorithmic transparency and explainability are examined, as well as the necessity for clarity regarding input and output data when artificial intelligence systems are used for evidentiary purposes. Through this comprehensive view on regulation, data logic, and the impact of automation, the section justifies the need for additional procedural rules that reflect the realities of the digital age.

### 4.1. The Data Act as a legal basis for procedural transformation

The adoption of Regulation (EU) 2023/2854, known as the Data Act [1], marks an important step in developing the legal framework for the digital economy in the European Union. Although the Act is not directly focused on procedural law, its provisions have a clear and direct impact on the ability of parties in judicial proceedings to access digital information, as well as on how this information is used, verified, and evaluated by courts. It introduces obligations for data holders, such as digital service providers or manufacturers of connected devices, to grant access to such data under certain conditions. These provisions are not procedural, yet objectively create conditions under which parties in judicial proceedings can request access to relevant digital information and build their evidentiary strategies based on it. This implies that courts will need to consider the right of access to data not only as a matter of substantive law but also as an element of procedural guarantees for a fair trial and effective defense.

This is the reason why the Data Act [1] serves as a legal instrument promoting the reform of the evidentiary framework. This requires the court to be prepared to evaluate digital datasets not only based on formal criteria but also regarding their structure, completeness, volume, and context, while simultaneously ensuring the lawful collection of such evidence during proceedings, whenever appropriate.

### 4.2. Data as evidence: from isolated units to behavioural patterns

Traditionally, in continental procedural law, evidence is perceived as separate, individualized units: written documents, expert reports, witness testimonies, or specific digital recordings. Historically, this approach has aligned with the principle of formal evidentiary value, enabling the assessment of each piece of evidence individually. In the context of digital transformation, this model proves insufficient for establishing objective truth. In this regard, conclusions have been drawn in various publications, including, but not limited to, those in the publication "Electronic Evidence in Criminal Cases and Their Sources—Necessary Reflections" [10]. Contemporary digital realities not only increase the volume of available information but also fundamentally alter the way this information is generated, organized, and utilized. In this context, further conclusions have been presented in various publications, such as Vidolov, I., B. Cherneva, Electronic Evidence and Electronic Evidence Means in Bulgarian Criminal Procedure, Globalization, the State and the Individual, No. 2(18), 2018, 35–44 [11], and Tsvetkova, Al., "Electronic Evidence in Bulgaria—A Step Forward, a Step Back," Society and Law, No. 8, 2018, 3–18 [12].

In judicial practice, there are increasingly frequent cases where a single informational unit is insufficient to resolve the legal issue at hand. Such an informational unit cannot be adequately assessed solely based on its formal evidentiary value, particularly when dealing with unstructured data. The behavior of a given subject—whether in the context of a phishing attack, breach of trust, or electoral violations—is not revealed through a single action, but rather through the repetition, chronology, and dynamics of actions over time. In this sense, data represent behavioral traces that can only be interpreted through a comprehensive analysis of the relationships between individual actions and various material media.

Shifting the perspective from "evidence as an informational unit" towards "evidence as a behavioral model" requires not only technological tools but also a methodological shift and adequate preparation for the collection and evaluation of aggregated digital evidence. This includes analyzing temporal sequences, assessing contextual credibility, verifying objectified information across different media types, and analyzing behavioral consistency. It also creates the opportunity to challenge not only an individual fact but also the underlying logic of the derived model, as well as its sufficiency and relevance.

Although courts already examine behavior within the broader context of proceedings, the lack of clear definitions, expert preparedness, and explicit procedural rules for dealing with behavioral models creates legal uncertainty for the parties, the court, and the appointed experts.

For this reason, we argue that the transition to an integrated, behavioral analysis of data should not be left solely to ad hoc expert assessment. There is an urgent need for clear regulation of admissibility, systematization, verification, and due analysis. The right of defense, in the context of analyzing datasets reflecting models rather than isolated events, should be realized through adequate procedural safeguards. The conclusion is that minimal procedural guarantees must be established for working with datasets, which, when viewed in their entirety, serve to reveal the objective truth.

### 4.3. Artificial Intelligence and judicial oversight of automated analyses

The growing integration of Artificial Intelligence (AI) into various spheres of public life presents the judicial system with fundamentally new and previously unknown challenges. One of the main questions is whether and to what extent courts can exercise control over the “decisions” made or prepared by AI systems, in order to ensure their legality.

The legal debate has already moved beyond the simple admissibility of digital evidence. It now encompasses additional issues such as the explainability of algorithms, access to input data, the controllability of the processing workflow, and the possibility to challenge the results of automated analysis. These concerns are also reflected in the Artificial Intelligence Act [4], which introduces requirements for transparency and accountability for high-risk AI systems, particularly in sensitive areas such as justice and law enforcement. The topic is also addressed by the Council of Europe in the Feasibility Study on a Legal Framework on Artificial Intelligence Design, Development and Application based on Council of Europe's Standards, 2021. Available at: https://rm.coe.int/cahai-2020-23-final-eng-feasibility-study-/1680a0c6da [13].

The general framework assumes that oversight, including judicial review, must be possible with respect to assessing whether a given algorithm is reliable and validated, as well as concerning the methodology and training data used. In the context of data, this also includes assessing whether the input data is complete, relevant, and accessible, and whether there is the possibility for independent expert verification, enabling the methodology to be reversed from output back to input or allowing for alternative interpretation. It is particularly important to emphasize that automated analyses are not evidence in themselves, but tools for extracting evidentiary value from data. Therefore, a clear distinction must be made between the factual basis (the data) and the interpretative model (the algorithm). Courts must ensure that both are subject to scrutiny and procedural challenge. They must be capable, both technically and methodologically, of carrying out such an analysis if the use of AI is to be regulated in accordance with the law.

It is widely accepted in the understanding of legality that AI must not operate as a "black box" whose results are accepted uncritically. In the context of a fair trial, both parties must have equal access to information about how a given piece of evidence has been generated or processed by an AI system. Only when technical explainability, legal admissibility, and the right to challenge are present, can legal relations involving AI systems be adequately protected procedurally.

## 5. Practical and institutional challenges in the analysis of data in judicial proceedings

The various legal relationships arising from the use of datasets, examined in the previous section within the context of their general regulatory framework, in our view highlight the need for courts to adopt a new model for working with digital evidence represented by datasets. Such datasets must be analyzed not as isolated units but as comprehensive behavioral models derived from the data. Although this approach is conceptually well-grounded, it cannot be practically implemented without adequate institutional, procedural, technical, and expert support. This section aims to identify the main barriers, including regulatory and infrastructural deficiencies, that limit the effective use of datasets as aggregated evidence in modern judicial proceedings.

First, the study highlights the lack of a clearly defined procedure for presenting and verifying digital datasets, especially when they consist of unstructured data, materialized on different media, and in some cases created or used across different time periods. In practice, courts are often presented with isolated excerpts, summary charts, or fragments of log files, without full disclosure or transparency regarding the selection process. Parties lack a formal procedural mechanism to request the complete dataset or to challenge the selectivity of the presented evidence. This creates a risk of contextual manipulation and impedes the construction of a realistic behavioral picture.

The current procedural framework does not account for the need for reverse expert analysis of models to verify the correctness of conclusions derived from input data. In disputes involving behavioral inferences drawn from data analysis, there is often no clarity about how the logic of the analysis itself can be contested, not just the numerical values. To ensure equality of arms, there must be a procedural opportunity for alternative expert interpretation of the same dataset, including the ability to present methodological counter-arguments.

When datasets are used as evidence in court, this requires a reliable system for the secure storage, review, and processing of large volumes of digital material. Expert capacity is also essential, including basic interdisciplinary competence among legal professionals who work with digital evidence and require an understanding of its technological characteristics. At present, lawyers are not trained to work with digital data sources (e.g., log files, metadata, video recordings, behavioral algorithms) and lack the methodological tools necessary to evaluate analytical models. This makes them heavily dependent on expert conclusions and unable to meaningfully challenge those conclusions in cases of disagreement. Targeted efforts are needed to build multidisciplinary expertise that combines knowledge from areas such as digital forensics, artificial intelligence, information systems, and data analysis..

It should be noted that the lack of procedural mechanisms for protecting the rights of the parties when working with datasets may compromise the proper exercise of the right to defense. A party that has no access to the algorithm or to the complete datasets cannot effectively challenge the expert's conclusions. This leads to a potential vulnerability in the right to defense and underscores the need for new procedural safeguards, such as the right of access to the full dataset, the right to verify the methodology in its entirety, and the right to object to the selection or aggregation of data or other technological aspects. Similar conclusions are argued in the publication by Ramos-Maqueda, M., D. L. Chen, The Data Revolution in Justice, World Development, 2024. https://doi.org/10.1016/j.worlddev.2024.106834 [14], as well as in Assoc. Prof. Dr. Dragomir Krastev, Use of Digital Evidence in Cybercrime Investigations, 2021, Higher School of Telecommunications and Posts, https://doi.org/10.36997/LBCS2021.332 [15].

In conclusion, it should be emphasized that the present section should be understood in conjunction with Section 4.2 of this study. While Section 4.2 outlines the theoretical need for a shift in the understanding of evidence, this section reveals the practical dimensions of that transformation. Without a clear procedural framework, adequate technical infrastructure, and trained legal professionals and experts, the transition toward analyzing data as behavioral models cannot be objectively implemented. It is precisely for this reason that reforms in this area must be directed not only at the conceptual foundations but also at the specific practical conditions necessary to ensure a fair and balanced process.

## 6. Conclusions

This study has examined both the legal and regulatory framework and the practical procedural challenges related to the use of data from information systems as evidence in judicial proceedings. Central to the analysis is the argument that it is necessary to fundamentally rethink how courts perceive and process digital evidence, not as isolated informational units, but as behavioural patterns extracted from comprehensive datasets. The conducted analysis leads to the following key conclusions below.

The new regulatory framework (the Data Act) establishes legal grounds for access to, use, and sharing of data, including within the judicial context. Although it is not procedural in form, it has a direct impact on the right to a defence and the equality of arms between the parties.

Digital evidence increasingly serves to reveal behavioural patterns, as it reflects complex factual circumstances rather than discrete facts. This necessitates a transition from fragmentary to systemic analysis and calls for a shift in the procedural logic of evidence presentation and verification.

Judicial practice already includes cases where datasets comprising logs, recordings, machine memory, and digital traces stored on various media are used. However, there is no unified methodology for interpreting and verifying such data, which creates risks of inconsistency and unequal treatment in similar cases.

Objective truth is often revealed through sequences of actions. When such actions occur in an electronic environment, there is a need to establish and apply new procedural standards and guarantees, such as: Full traceability and access to the dataset, including audio recordings, machine logs, video files, and metadata; Expert analysis focused on behavioural interpretation rather than purely technical or linguistic evaluation; The opportunity for parties to participate in data analysis; Judicial oversight over the selection of data submitted as evidence, particularly when it is a sample from a broader communication or digital context; Recognition of behavioural patterns and repetitions, not just isolated phrases or actions.

Courts must play an active role as guarantors of procedural rights, particularly when dealing with the specificities of datasets. This includes ensuring the right of access to the full dataset, the right to challenge selective samples, and the right to review algorithmic analyses.

There is a lack of sufficiently trained legal professionals and institutional infrastructure capable of ensuring reliable handling of datasets. This impairs the judiciary's ability to reach reasoned decisions based on complex digital analysis.

In conclusion, the use of data as evidence in contemporary judicial proceedings is an inevitable consequence of society's digital transformation. To ensure that this process is fair, transparent, and effective, a shift is required not only in legal reasoning but also in the institutional preparedness of the judiciary. This transformation is not solely a matter of technology—it is a matter of justice keeping pace with technological advancement in order to uphold the protection of citizens' rights and the discovery of objective truth.

*Dobromira Bankova and Vladimir Dimitrov*
*Faculty of Mathematics and Informatics*
*Sofia University "St. Kliment Ohridski"*
*5 James Bourchier Blvd.*
*1164 Sofia*
*BULGARIA*
*E-mail: cht@fmi.uni-sofia.bg*